\documentclass[pre,twocolumn,superscriptaddress,nofootinbib]{revtex4}

\usepackage{graphicx,chemarr,color}

\begin{document}

\title{Universality of biochemical feedback and its application to immune cells}

\author{Amir Erez}
\thanks{These authors contributed equally.}
\affiliation{Department of Molecular Biology, Princeton University, Princeton, NJ 08544, USA}

\author{Tommy A.\ Byrd}
\thanks{These authors contributed equally.}
\affiliation{Department of Physics and Astronomy, Purdue University, West Lafayette, Indiana 47907, USA}

\author{Robert M.\ Vogel}
\affiliation{IBM T.\ J.\ Watson Research Center, Yorktown Heights, New York 10598, USA}

\author{Gr\'egoire Altan-Bonnet}
\affiliation{Immunodynamics Group, Cancer and Inflammation Program, National Cancer Institute, National Institutes of Health, Bethesda, Maryland 20814, USA}

\author{Andrew Mugler}
\email{amugler@purdue.edu}
\affiliation{Department of Physics and Astronomy, Purdue University, West Lafayette, Indiana 47907, USA}

\begin{abstract}
We map a class of well-mixed stochastic models of biochemical feedback in steady state to the mean-field Ising model near the critical point. The mapping provides an effective temperature, magnetic field, order parameter, and heat capacity that can be extracted from biological data without fitting or knowledge of the underlying molecular details. We demonstrate this procedure on fluorescence data from mouse T cells, which reveals distinctions between how the cells respond to different drugs. We also show that the heat capacity allows inference of absolute molecule number from fluorescence intensity. We explain this result in terms of the underlying fluctuations and demonstrate the generality of our work.
\end{abstract}

\maketitle

\section{Introduction}

Positive feedback is ubiquitous in biochemical networks and can lead to a bifurcation from a monostable to a bistable cellular state \cite{mitrophanov2008positive, tkavcik2012optimizing, das2009digital, vogel2016dichotomy}. Near the bifurcation point, the bistable state often reflects a choice between two accessible but opposing cell fates. For example, in T cells, the distribution of doubly phosphorylated ERK (ppERK) can be bimodal \cite{vogel2016dichotomy}. ppERK is a protein that initiates cell proliferation and is implicated in the self/non-self decision between mounting an immune response or not \cite{vogel2016dichotomy, altan2005modeling}.

The bifurcation point is similar to an Ising-type critical point in physical systems such as fluids, magnets, and superconductors, where a disordered state transitions to one of two ordered states at a critical temperature \cite{goldenfeld1992lectures}. In fact, universality tells us that the two should not just be similar, they should be the same: because they are both bifurcating systems, both types of systems should exhibit the same critical scaling exponents and therefore belong to the same universality class \cite{goldenfeld1992lectures}. Although this powerful idea has allowed diverse physical phenomena to be united into specific behavioral classes, the application of universality to biological systems is still developing \cite{mora2011biological, munoz2018colloquium, salman2012universal, brenner2015universal, pal2014non, ridden2015entropy, qian2016framework, hidalgo2014information}.

Biological tools such as flow cytometry, fluorescence microscopy, and RNA sequencing allow reliable experimental estimates of abundance distributions, inspiring researchers to seek to apply insights from statistical physics to biological data. In particular, recent studies have demonstrated that biological systems on many scales, from molecules \cite{mora2010maximum}, to cells \cite{kastner2015critical, krotov2014morphogenesis, de2017critical, chen2012scale, aguilar2018critical, wan2018time}, to populations \cite{bialek2014social, attanasi2014finite, cavagna2017dynamic}, exhibit signatures consistent with physical systems near a critical point. However, some of these studies have come under scrutiny because some of the signatures, particularly scaling laws, can arise far from or independent of a critical point \cite{schwab2014zipf, touboul2017power, newman2005power}. Part of the problem is that the identification of appropriate scaling variables from data can be ambiguous, and one is often left looking for scaling relationships in an unguided way. 

Typical approaches to the interpretation of abundance distributions include fitting to either detailed mechanistic models of the underlying reaction scheme, or to an effective description of the data such as a Gaussian or lognormal mixture model. The former approach is usually difficult to parameterize and difficult to generalize to other systems. The latter approach often suffers from numerical issues (the likelihood is unbounded and the expectation-maximization algorithm can lead to spurious solutions \cite{biernacki2003choosing}). Moreover, the vicinity of a bifurcation point is precisely where a mixture analysis is most likely to fail. In contrast, mapping to a statistical physics framework is expected to be universal, in the sense that the precise microscopic details of a broad range of biochemical models are unimportant near the bifurcation point, as they are coarse-grained rather than particular reaction parameters.

Here we provide a framework for mapping well-mixed stochastic models of biochemical feedback to the mean-field Ising model and apply it to published data on T cells. This allows us to extract effective thermodynamic quantities from experimental data without needing to fit to a parametric model of the system. This makes the theory applicable to a broad class of biological datasets without worrying about model selection or goodness-of-fit criteria. The theory provides insights on how T cells respond to drugs and reveals distinctions between one type of drug response and another. Furthermore, we find that one of the thermodynamic quantities (the heat capacity) provides a novel way to estimate absolute molecule number from fluorescence level in bifurcating systems. We demonstrate that our results can be extended to cases where feedback is indirect and discuss further extensions, including to spatiotemporal dynamics.

\section{Results}

\begin{figure}[t]
\centering
\includegraphics[width=\linewidth]{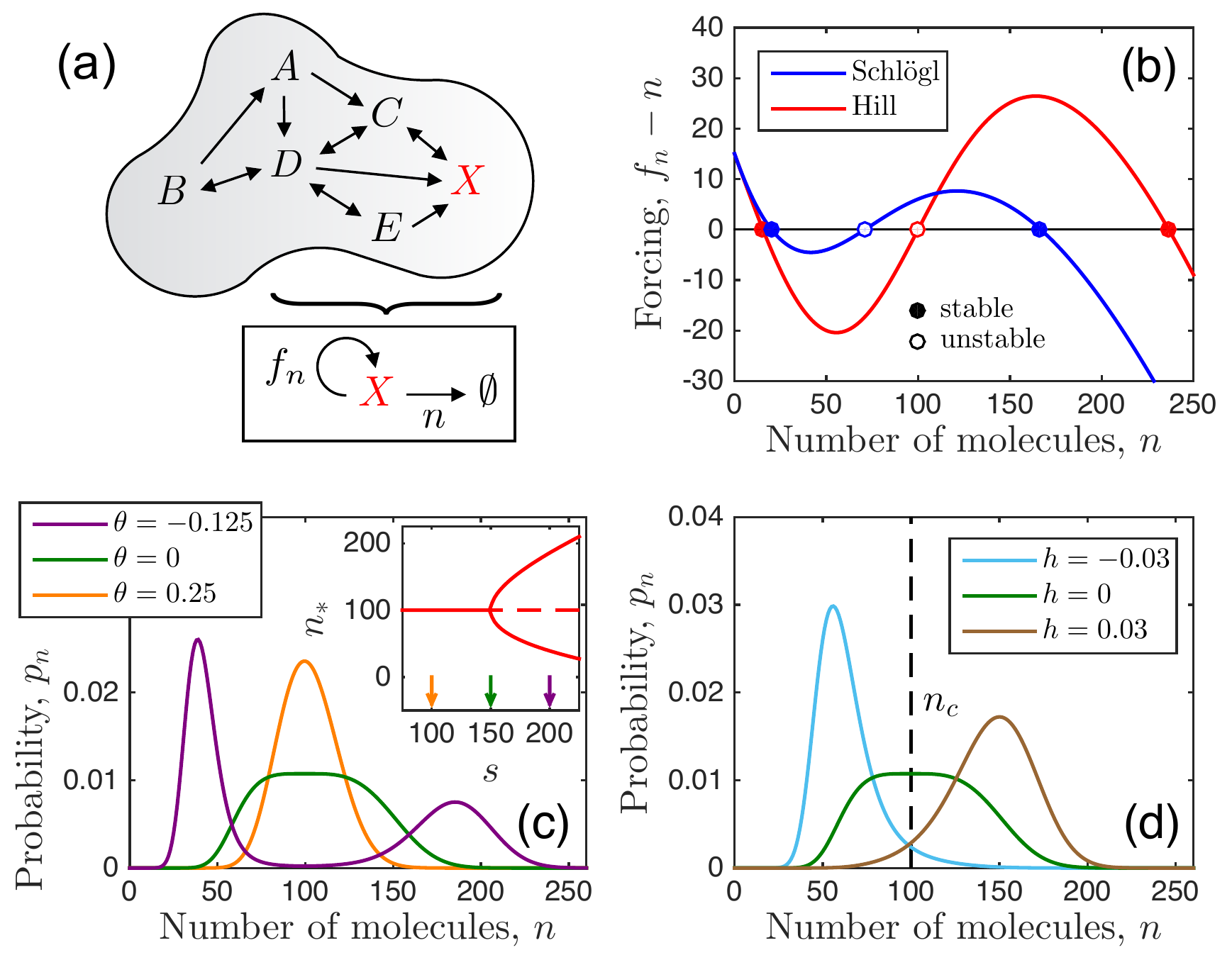}
\caption{Setup and behavior of the model. (a) We consider well-mixed stochastic biochemical networks described by an effective feedback function $f_n$. (b) Feedback produces either one or two stable steady states. (c) The molecule number distribution is peaked around these states or flat at the bifurcation point. (d) Mapping to the Ising model reveals that the effective reduced temperature drives the distribution to the unimodal ($\theta > 0$) or bimodal ($\theta < 0$) state (see c), while the effective field $h$ biases the distribution toward high ($h > 0$) or low ($h < 0$) molecule number. Parameters: $H=3$ and $n_c = 100$ in b, c, and d; $h = 0$ in b and c; and $\theta = 0$ in d (see also Appendix \ref{app:param}).}
\label{fig:setup}
\end{figure}

We consider a reaction network in a cell where $X$ is the molecular species of interest, and the other species $A$, $B$, $C$, etc.\ form a chemical bath for $X$ [Fig.\ \ref{fig:setup}(a)]. The reactions of interest produce or degrade an $X$ molecule, can involve the bath species, and in principle are reversible. We allow for nonlinear feedback on $X$, meaning that the production of an $X$ molecule in a particular reaction might require a certain number of $X$ molecules as reactants. This leads to an arbitrary number of reactions of the form
\begin{equation}
\label{eq:rxns}
j_rX + Y_r^+ \xrightleftharpoons[k_r^-]{k_r^+} (j_r+1)X + Y_r^-,
\end{equation}
where in the $r$th reaction, $j_r$ are stoichiometric integers describing the nonlinearity, $k_r^\pm$ are the forward ($+$) and backward ($-$) reaction rates, and $Y_r^\pm$ represent bath species involved as reactants ($+$) or products ($-$). A simple and well-studied special case of Eq.\ \ref{eq:rxns} is Schl\"ogl's second model \cite{schlogl1972chemical, dewel1977renormalization, nicolis1980systematic, brachet1981critical, grassberger1982phase, prakash1997dynamics, liu2007quadratic, vellela2009stochastic}, in which $X$ is either produced spontaneously from bath species $A$, or in a trimolecular reaction from two existing $X$ molecules and bath species $B$ (i.e., $R = 2$, $j_1 = 0$, $j_2 = 2$, $Y_1^+ = A$, $Y_2^+ = B$, and $Y_1^- = Y_2^- = \emptyset$).

We assume that molecules are well-mixed and that the numbers of bath molecules are constant. The latter assumption is equivalent to integrating out all species but $X$, such that the feedback on $X$ arises directly from $X$ itself (Eq.\ \ref{eq:rxns}). However, in general the feedback will be indirect, with $X$ regulating dynamic species in the bath that in turn regulate $X$ (this is almost certainly the case in the T cells we study here). Therefore, we consider this more general case later in Section \ref{sec:indirect} and show that the results discussed below remain unchanged.

The master equation for the probability of observing $n$ molecules of species $X$ according to Eq.\ \ref{eq:rxns} is
\begin{equation}
\label{eq:me}
\dot{p}_n = b_{n-1}p_{n-1} + d_{n+1}p_{n+1} - (b_n + d_n)p_n,
\end{equation}
where $b_n = \sum_{r=1}^R J_{rn}^+$ and $d_n = \sum_{r=1}^R J_{rn}^-$ are the total birth and death propensities, and $J_{rn}^+ = k_r^+ n_r^+ n!/(n-j_r)!$ and $J_{rn}^- = k_r^- n_r^- n!/(n-j_r-1)!$ are the forward and backward propensities of each reaction pair. Here $n_r^\pm$ are the numbers of molecules of the bath species involved in reaction $r$, and the factorials account for the number of ways that $X$ molecules can meet in a reaction. The steady state of Eq.\ \ref{eq:me} is \cite{van1992stochastic, gardiner1985handbook}
\begin{equation}
\label{eq:pn}
p_n = p_0 \prod_{j = 1}^n \frac{b_{j-1}}{d_j} = \frac{p_0}{n!} \prod_{j=1}^n f_j,
\end{equation}
where $p_0^{-1} = \sum_{n=0}^\infty(1/n!)\prod_{j=1}^n f_j$ is set by normalization. In the second step of Eq.\ \ref{eq:pn} we define an effective birth propensity $f_n \equiv nb_{n-1}/d_n$ corresponding to spontaneous death with propensity $n$ [Fig.\ \ref{fig:setup}(a)]. In general, $f_n$ is an arbitrary, nonlinear feedback function governed by the reaction network. For the Schl\"ogl model, it is $f_n = [aK^2 + s(n-1)(n-2)]/[(n-1)(n-2)+K^2]$, where we have introduced the dimensionless quantities $a \equiv k_1^+n_A/k_1^-$, $s \equiv k_2^+ n_B/k_2^-$, and $K^2 \equiv k_1^-/k_2^-$. As a ubiquitous example we also consider the Hill function $f_n = a + sn^H/(n^H+K^H)$ with coefficient $H$. Importantly, the inverse of Eq.\ \ref{eq:pn},
\begin{equation}
\label{eq:fn}
f_n = \frac{np_n}{p_{n-1}},
\end{equation}
allows calculation of the feedback function from the distribution \cite{walczak2009stochastic}, as utilized when analyzing the experimental data later in Section \ref{sec:immune}.

The quantity $f_n-n$ determines the dynamic stability: there can be either one or two stable states $n_*$ [Fig.\ \ref{fig:setup}(b)], and the transition from a monostable to a bistable regime occurs at a bifurcation point [Fig.\ \ref{fig:setup}(c) inset]. These deterministic regimes correspond stochastically to unimodal and bimodal distributions $p_n$, respectively, with maxima at $n_*$, while the bifurcation point corresponds to a distribution that is flat on top [Fig.\ \ref{fig:setup}(c)].

\subsection{Ising mapping and scaling exponents}

To understand the scaling behavior near the bifurcation point, we expand the stability condition $f_{n_*}-n_*=0$ to third order around a point $n_c$ satisfying $f''_{n_c} = 0$. This choice of $n_c$ eliminates the quadratic term in the dynamic forcing $f_n-n$, equivalent to eliminating the cubic term in an effective potential as in Ginzburg--Landau theory \cite{kopietz2010introduction}. Defining the parameters
\begin{equation}
\label{eq:cparam}
m \equiv \frac{n_*-n_c}{n_c}, \quad
h \equiv \frac{2(f_{n_c} - n_c)}{-f'''_{n_c}n_c^3}, \quad
\theta \equiv \frac{2(1-f'_{n_c})}{-f'''_{n_c}n_c^2},
\end{equation}
the expansion $f_{n_c} + f'_{n_c}(n_*-n_c) + f'''_{n_c}(n_*-n_c)^3/3! - n_* = 0$ becomes $h - \theta m - m^3/3 = 0$. This expression is equivalent to the expansion of the Ising mean field equation $m = \tanh[(m+h)/(1+\theta)]$ for small magnetization $m$, where $\theta = (T-T_c)/T_c$ is the reduced temperature, and $h$ is the dimensionless magnetic field \cite{kopietz2010introduction}. Therefore, in our system we interpret $m$ as the order parameter, $\theta$ as an effective reduced temperature, and $h$ as an effective field. Explicit expressions for $n_c$, $\theta$, and $h$ in terms of the  biochemical parameters and vice versa are given for the Schl\"ogl and Hill models in Appendix \ref{app:param}.

We see in Fig.\ \ref{fig:setup}(c) and (d) that $n_c$ determines where the distribution is centered, that $\theta$ drives the system to the unimodal ($\theta > 0$) or bimodal ($\theta < 0$) state, and that $h$ biases the system to high ($h > 0$) or low ($h < 0$) molecule numbers. Note that unlike in the Ising model, even when $h=0$ an asymmetry persists between the high and low states [see the purple distribution in Fig.\ \ref{fig:setup}(c)]. The reason is that in the master equation (Eq.\ \ref{eq:me}), unlike in Ginzburg--Landau theory, fluctuations scale with molecule number, such that the high state is wider than the low state.

The equivalence between our system and the Ising mean-field equation near the critical point (Eq.\ \ref{eq:cparam}) implies that our system has the same scaling exponents $\beta=1/2$, $\gamma=1$, and $\delta=3$ as the Ising universality class in its mean-field limit \cite{kopietz2010introduction}. For completeness, we verify in Appendix \ref{app:scalings} that these scalings are indeed obeyed by the Schl\"ogl and Hill models.

\begin{figure}[t]
\centering
\includegraphics[width=\linewidth]{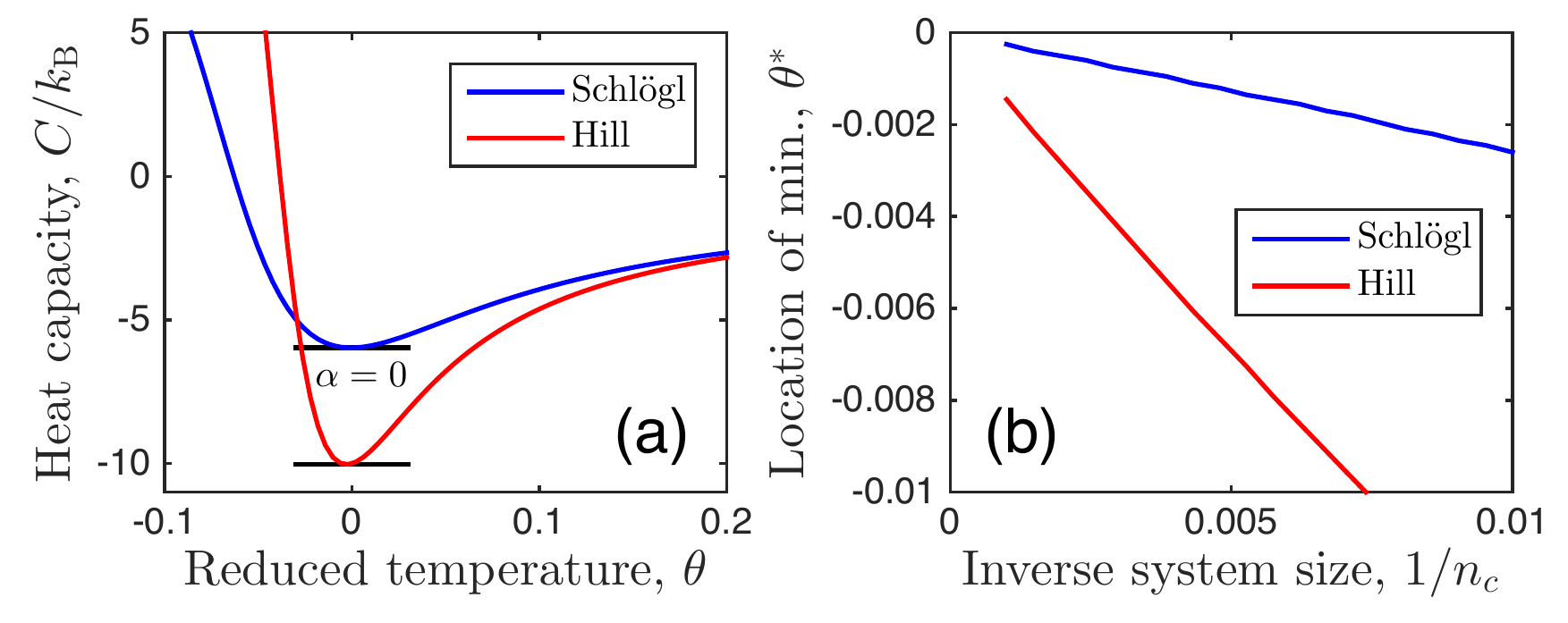}
\caption{(a) Heat capacity (Eq.\ \ref{eq:C}) is minimized at the bifurcation point, corresponding to exponent $\alpha = 0$. (b) The location of the minimum approaches $\theta^*\to0$ as $n_c\to\infty$, as expected. Parameters: $H=3$, $n_c = 500$, and $h=0$.}
\label{fig:heat1}
\end{figure}

However, Eq.\ \ref{eq:cparam} does not explicitly determine the value of the exponent $\alpha$. The reason is that, unlike $\beta$, $\gamma$, and $\delta$, the exponent $\alpha$ depends on the entire distribution $p_n$, not just the maxima.  Specifically, $\alpha$ concerns the heat capacity, $C|_{h=0}\sim |\theta|^{-\alpha}$, which depends on the entropy $S$ and thus $p_n$. The equilibrium definition $C = T\partial_T S$ generalizes to a nonequilibrium system like ours when one uses the Shannon entropy $S = -k_{\rm B}\sum_n p_n \log p_n$ \cite{mandal2013nonequilibrium}. Since $T = (1+\theta)T_c$, we have $C = (1+\theta)\partial_\theta S$, or
\begin{equation}
\label{eq:C}
\frac{C}{k_{\rm B}} = -(1+\theta)\sum_{n=0}^\infty p_n (1+\log p_n) \Bigg( \psi_n - \sum_{j=0}^\infty p_j \psi_j \Bigg),
\end{equation}
where $\psi_n \equiv (1/2)f'''_{n_c}n_c^2\sum_{j=1}^n(j-n_c)/f_j$. Eq.\ \ref{eq:C} follows from performing the $\theta$ derivative using the expression in Eq.\ \ref{eq:pn}, the expansion below Eq.\ \ref{eq:cparam}, and the definition of $\theta$ (Eq.\ \ref{eq:cparam}). We see in Fig.\ \ref{fig:heat1}(a) that when $h=0$, $C$ exhibits a minimum at $\theta^*$. We see in Fig.\ \ref{fig:heat1}(b) that $\theta^*$ vanishes as the system size increases, $n_c\to\infty$. This implies that $C|_{h=0}\sim |\theta|^0$ to sub-quadratic order in $\theta$, or $\alpha = 0$, again consistent with the Ising universality class in its mean-field limit. Interestingly, whereas $C$ is discontinuous in the mean-field Ising model \cite{kopietz2010introduction} and constant in the van der Waals model of a fluid \cite{goldenfeld1992lectures}, it is minimized here; nevertheless, in all cases $\alpha = 0$. Note from Fig.\ \ref{fig:heat1}(a) that $C$ is negative near $\theta = 0$; negative heat capacity is a well-known feature of nonequilibrium steady states \cite{zia2002getting, boksenbojm2011heat, bisquert2005master}.

\subsection{Application to immune cell data}
\label{sec:immune}

\begin{figure}[t]
\centering
\includegraphics[width=\linewidth]{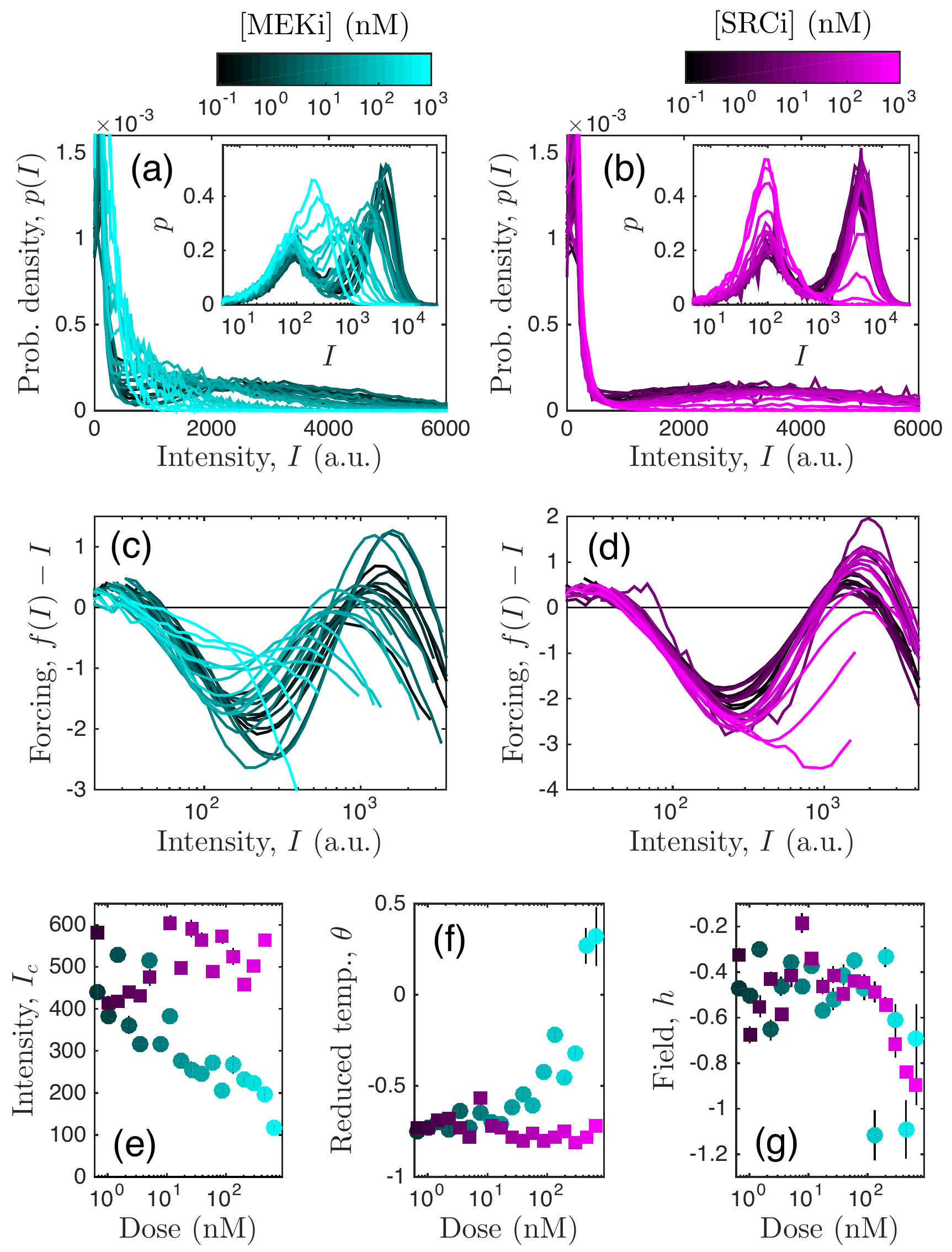}
\caption{Application of the theory to immune cell data. Upon administration of either (a) MEK or (b) SRC inhibitor, experimental distributions of T cell ppERK fluorescence intensity are unimodal (bimodal) for high (low) doses. Insets show distributions of log intensity for clarity. (c, d) Feedback functions calculated from the experimental distributions correspondingly exhibit either one or two stable states. (e-g) Effective thermodynamic quantities calculated from the data vary with drug dose in distinct ways for each drug. The results in (c-g) corroborate those in \cite{vogel2016dichotomy}, but with a much simpler framework that has three parameters instead of five and requires no fitting or prior biological knowledge of the system. Error bars: standard error from filter windows $25 \le W \le 35$ (see Appendix \ref{app:analysis}).}
\label{fig:experiment}
\end{figure}

To demonstrate the utility of our theory, we apply it to published data from T cells \cite{vogel2016dichotomy}. In these experiments, chemotherapy drugs inhibit the enzymes MEK and SRC in the biochemical networks of the cells. The inhibition results in bimodal (low dose) or unimodal (high dose) distributions of ppERK abundance, which is measured as fluorescence intensity $I$ by flow cytometry. The distributions are shown for a range of drug doses in Fig.\ \ref{fig:experiment}(a) and (b) (the insets show distributions of log intensity for clarity). Experimental details are given in the original publication \cite{vogel2016dichotomy} and are summarized in Appendix \ref{app:expt}, along with the drugs and dose amounts.

First, we compute the feedback function $f$ from each distribution using Eq.\ \ref{eq:fn} (see Appendix \ref{app:analysis}). Fig.\ \ref{fig:experiment}(c) and (d) show the corresponding forcing functions [compare to Fig.\ \ref{fig:setup}(b)]. As expected, in each case we see that the forcing function transitions from two stable states to one stable state as the drug is applied.

Then, we compute $I_c$ (the analog of $n_c$ in units of fluorescence intensity), $\theta$, and $h$ from the feedback function using Eq.\ \ref{eq:cparam} (see Appendix \ref{app:analysis}). These quantities are shown as a function of drug dose in Fig.\ \ref{fig:experiment}(e)-(g). We see that the behavior is different depending on whether MEK inhibitor (MEKi) or SRC inhibitor (SRCi) is applied. Specifically, MEKi decreases $I_c$, increases $\theta$, and decreases $h$; whereas SRCi only decreases $h$, leaving the other quantities unchanged. Thus, the effective thermodynamic quantities can differentiate cellular responses to different perturbations, such as the application of different drugs.

Furthermore, the mapping provides an intuitive interpretation of the drug responses. MEKi causes a transition from a bimodal to a unimodal state in the expected way: by increasing the reduced temperature $\theta$ from a negative to a positive value [Fig.\ \ref{fig:experiment}(f)]. In the process, $I_c$ decreases [Fig.\ \ref{fig:experiment}(e)], meaning that the unimodal state is shifted to lower molecule number, near the lower mode of the bimodal state [Fig.\ \ref{fig:experiment}(a) inset]. In contrast, SRCi causes a transition from a bimodal to a unimodal state in a different way: by decreasing the field while leaving $\theta$ and $I_c$ unchanged [Fig.\ \ref{fig:experiment}(e)-(g)]. In essence, the distribution remains bimodal and unshifted, except that the field causes the high mode to diminish in weight [Fig.\ \ref{fig:experiment}(b) inset]. Interestingly, the mean dose-response curves are similar for the two drugs \cite{vogel2016dichotomy}, but our mapping elucidates precisely how the transitions are different at the distribution level. Related conclusions were drawn in \cite{vogel2016dichotomy}, but those conclusions relied on fitting the distributions to a five-parameter Gaussian mixture model, which is expected to fail near the bifurcation point. Here we use only three parameters and no fitting, and we emerge with an intuitive interpretation in terms of thermodynamic quantities.

Finally, we note that for both drugs the effective field is negative at all doses [Fig.\ \ref{fig:experiment}(g)]. The reason is that the fluorescence distributions have long tails (which is why they are often easier to visualize in log space); see Fig.\ \ref{fig:experiment}(a) and (b). In the theory, a long tail is indistinguishable from a low-molecule-number bias in the peak, which corresponds to $h < 0$. We address the possible origins and implications of the long tails in the Discussion (Section \ref{sec:discussion}).

\subsection{Estimation of molecule number}

\begin{figure}[t]
\centering
\includegraphics[width=\linewidth]{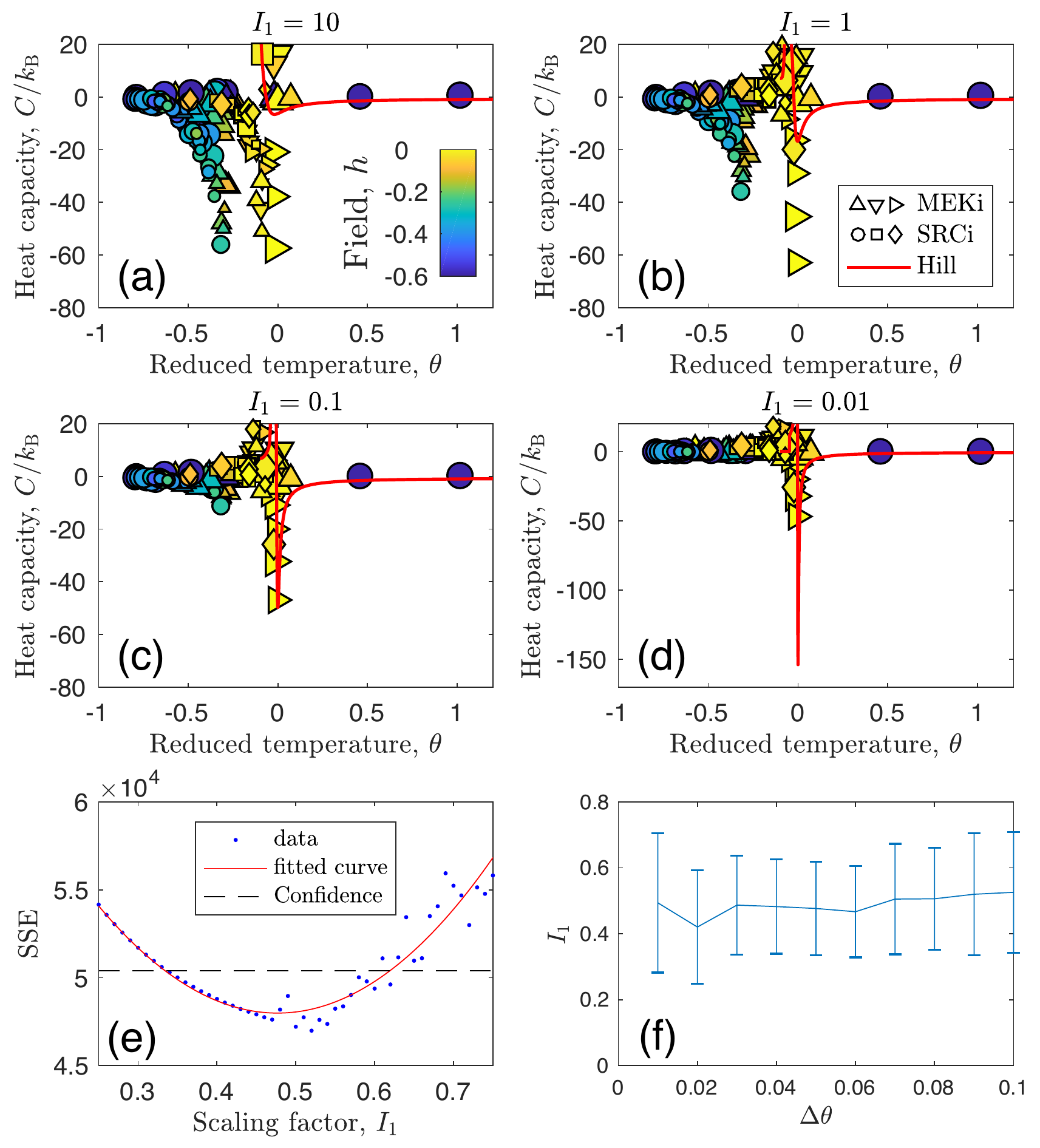}
\caption{Estimation of molecule number by comparing heat capacity between theory and experiments. (a-d) Rough estimate of fluorescence-to-molecule-number conversion factor $I_1$ (see titles) obtained by comparing depths of theory and experimental minima. ``Hill'' refers to the theoretical curve produced by Hill-function feedback as in Eq.~\ref{eq:Hill}. Different symbols correspond to different drugs. See Appendix \ref{app:expt} for drugs (shape) and doses (size). (e) More precise estimate obtained from plotting sum of squared errors (SSE) for data within $-\Delta\theta \le \theta \le \Delta\theta$ and fitting to parabola (see Appendix D for details). Here $\Delta\theta = 0.05$. (f) Estimate is insensitive to value of $\Delta\theta$. Theory parameters: $H = 4$, $h = 0$, and $n_c = \bar{I}_c/I_1$, where $\bar{I}_c = 730$ is the average value across all experiments.}
\label{fig:heat2}
\end{figure}

We now apply the theory to compute the heat capacity from the T cell data.
Specifically, we compute $C$ using Eq.\ \ref{eq:C} (see Appendix \ref{app:analysis}) for all drugs and doses used in the experiments \cite{vogel2016dichotomy} (Appendix \ref{app:expt}). Unlike the other thermodynamic quantities, $C$ requires a conversion from fluorescence intensity to molecule number because it depends explicitly on the distribution $p_n$ (Eq.\ \ref{eq:C}). Therefore we compute $C$ for various values of the conversion factor $I_1$, where $n = I/I_1$. The results are shown in Fig.\ \ref{fig:heat2}. We see that irrespective of $I_1$ over four orders of magnitude, the data closest to $h=0$ (yellow) exhibit a global minimum in $C$ at $\theta = 0$, as expected from Fig.\ \ref{fig:heat1}(a). However, we also see that the depth of the minimum agrees with that of the theory only for the particular choice $I_1 \approx 0.1$ [Fig.\ \ref{fig:heat2}(c)].

To obtain a more precise estimate of $I_1$, we plot the sum of squared errors between the data and the theory as a function of $I_1$ in Fig.\ \ref{fig:heat2}(e). We focus on the bifurcation region by considering only values of $\theta$ within $-\Delta\theta \le \theta \le \Delta\theta$, and we find that our results are not sensitive to the choice of $\Delta\theta$ [Fig.\ \ref{fig:heat2}(f)]. This procedure (see the details in Appendix \ref{app:analysis}) results in an estimate of $I_1 = 0.5$ $\pm$ $0.2$, as seen in Fig.\ \ref{fig:heat2}(f). This value of $I_1$ corresponds to $\bar{n}_*=$ 170,000 $\pm$ 70,000 ppERK molecules in the high mode averaged across all cases with no inhibitor. It is possible to compare this value with previous measurements on these cells. In two separate experiments, it was estimated that there are approximately 100,000  \cite{altan2005modeling} and 214,000 \cite{hukelmann2016cytotoxic} ERK molecules per cell, and that only about 50\% of these molecules are doubly phosphorylated during T cell receptor activation \cite{altan2005modeling} (see Appendix \ref{app:analysis}). These considerations give a range of roughly 50,000$-$107,000 ppERK molecules, which is consistent with our estimate of 170,000 $\pm$ 70,000. The agreement is especially notable given that T cell protein abundances generally span six orders of magnitude, from tens to tens of millions of molecules per cell \cite{hukelmann2016cytotoxic}.

Why does the heat capacity extract the conversion between fluorescence intensity and molecule number? As mentioned above, $\alpha$ is the only exponent that is a function of $p_n$ instead of just its maxima. This means that the plot of $C$ vs.\ $\theta$ contains information not only about means or modes, but also about fluctuations. The notion that fluctuation information is essential for converting from intensity to molecule number can be seen with a simpler example: a Poisson distribution. Here we would have $\sigma_I^2/\bar{I}^2 = \sigma_n^2/\bar{n}^2 = 1/\bar{n} = I_1/\bar{I}$. From this relation it is clear that information about not only the mean ($\bar{I}$) but also the fluctuations ($\sigma_I^2$) in intensity is necessary and sufficient to infer the conversion factor $I_1$. In our case, the heat capacity is extracting similar information, but for a bifurcating system.


\subsection{Generalization to indirect feedback}
\label{sec:indirect}

In the T cells, it is well known that ppERK does not apply feedback to its own activation directly, but rather indirectly via upstream components \cite{vogel2016dichotomy, shin2009positive, altan2005modeling}. Therefore, we seek to determine the extent to which the above results are sensitive to our assumption in the theory that the feedback is direct. To this end, we construct a minimal extension of the model in Eq.\ \ref{eq:rxns} in which the feedback is indirect:
\begin{align}
& \emptyset \xrightleftharpoons[k_2]{k_1} X, \qquad
2X \xrightleftharpoons[k_4]{k_3} D, \nonumber \\
& D \xrightarrow{k_5} D + A, \qquad
A \xrightarrow{k_6} A + X, \qquad
A \xrightarrow{k_7} \emptyset, \nonumber \\
\label{eq:indirect}
& D \xrightarrow{k_8} D + B, \qquad
B + X \xrightarrow{k_9} B, \qquad
B \xrightarrow{k_{10}} \emptyset.
\end{align}
Here $X$ is produced, is degraded, and reversibly dimerizes (first line); the dimer $D$ produces a species $A$ that produces $X$ and is degraded (second line); and the dimer also produces a species $B$ that degrades $X$ and is degraded (third line). Eq.\ \ref{eq:indirect} is an extension of Eq.\ \ref{eq:rxns} because there are multiple stochastic variables ($X$, $D$, $A$, and $B$), there are irreversible reactions, and $X$ feeds back on itself indirectly through $D$, $A$, and $B$ instead of directly.

The deterministic steady state of Eq.\ \ref{eq:indirect} is
\begin{equation}
\label{eq:multidet}
0 = \dot{n}/k_2 = c_0 - n_* + c_2 n_*^2 - c_3 n_*^3,
\end{equation}
where $c_0 \equiv k_1/k_2$, $c_2 \equiv k_3k_5k_6/(k_2k_4k_7)$, $c_3 \equiv k_3k_8k_9/(k_2k_4k_{10})$, and the molecule numbers of $D$, $A$, and $B$ have been eliminated in favor of $n_*$ by setting their own time derivatives to zero. Because Eq.\ \ref{eq:multidet} is cubic in $n_*$, we see immediately that it has the same form as the expanded Ising mean field equation $h - \theta m - m^3/3 = 0$ (see Eq.\ \ref{eq:cparam}). Specifically, defining $m = (n_*-n_c)/n_c$ as in Eq.\ \ref{eq:cparam}, the choice $n_c = c_2/(3c_3)$ eliminates the term quadratic in $m$ and implies $\theta = 3c_3/c_2^2 - 1$ and $h = 9c_0c_3^2/c_2^3 - 3c_3/c_2^2 +2/3$. It immediately follows that this model has the same exponents  $\beta=1/2$, $\gamma=1$, and $\delta=3$ as the mean-field Ising universality class.

To test whether the heat capacity for this model exhibits the same features as that for the direct feedback model in Fig.\ \ref{fig:heat1}(a), we compute the steady state marginal distribution $p_n$ using stochastic simulations \cite{gillespie1977exact} of Eq.\ \ref{eq:indirect}. Specifically, we set $k_3/k_4 = 1/n_c$ and $k_5/k_7 = k_8/k_{10} = 1$ to ensure that the numbers of $D$, $A$, and $B$ molecules, respectively, are on the order of $n_c$. We then set $k_4/k_2 = k_7/k_2 = k_{10}/k_2 = \rho$, where $\rho$ is a free parameter that determines whether the degradation timescales of $D$, $A$, and $B$, respectively, are faster ($\rho > 1$) or slower ($\rho < 1$) than that of $X$. These conditions, along with the definitions of $n_c$, $\theta$, and $h$ above, constitute nine equations for nine reaction rates, plus $k_2$ which sets the units of time. Solving these equations yields expressions for the rates in terms of $n_c$, $\theta$, $h$, and $\rho$ that we use in the simulations.

\begin{figure}[t]
\centering
\includegraphics[width=\linewidth]{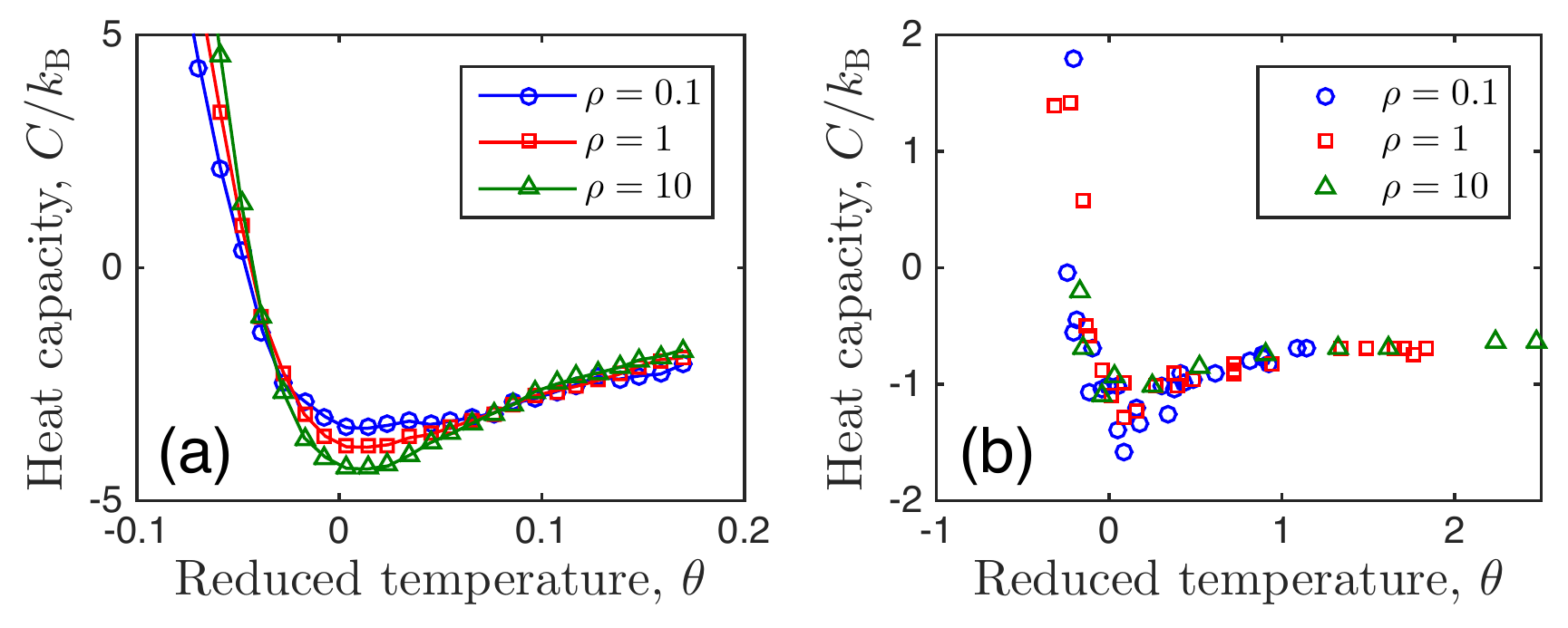}
\caption{Verification that indirect feedback does not qualitatively change modeling assumptions or results. (a) $C$ and $\theta$ calculated from extended model with indirect feedback. (b) $C$ and $\theta$ inferred assuming the feedback is direct (Eq.\ \ref{eq:fn}). Compare with Fig.\ \ref{fig:heat1}(a). Parameters: $n_c=100$ and $h=0$.}
\label{fig:indirect}
\end{figure}

Fig.\ \ref{fig:indirect}(a) shows the heat capacity $C$ as a function of $\theta$ for $h = 0$, $n_c = 100$, and $\rho = \{0.1, 1, 10\}$, where $C = (1+\theta)\partial_\theta S$ is computed from the entropy $S = -k_{\rm B}\sum_n p_n \log p_n$ by numerical derivative. We see that for all $\rho$ values, the curves exhibit a minimum at $\theta = 0$, implying $\alpha = 0$, and they rise more steeply for negative than for positive $\theta$ as in Fig.\ \ref{fig:heat1}(a).

We then investigate whether Eq.\ \ref{eq:rxns} remains valid as a coarse-grained description of the extended model in Eq.\ \ref{eq:indirect}. To answer this question, we infer values of $n_c$, $\theta$, $h$, and $C$ directly from the simulation data $p_n$ using the same protocol as for the experimental data. That is, we compute $f_n$ via Eq.\ \ref{eq:fn}, and then compute $\theta$, $h$, and $C$ from its derivatives at $n_c$ according to Eqs.\ \ref{eq:cparam} and \ref{eq:C}, where $n_c$ satisfies $f''_{n_c} = 0$. As with the experimental data (see Appendix \ref{app:analysis}), derivatives are calculated using a Savitsky-Golay filter \cite{savitzky1964smoothing}, although here we apply the filter directly to $f_n$ and perform the analysis directly in $n$ space, not log space.

Fig.\ \ref{fig:indirect}(b) shows the result of this procedure for the inferred heat capacity $C$ as a function of the inferred $\theta$. We see that, as with the exact $C$ and $\theta$ [Fig.\ \ref{fig:indirect}(a)], the data exhibit a minimum at $\theta = 0$ and rise more steeply for negative than for positive $\theta$. Note that the values of $C$ and $\theta$ are different in (a) and (b), which is expected because the shape of $p_n$ is not quantitatively the same in the two models of Eqs.\ \ref{eq:rxns} and \ref{eq:indirect}; nonetheless, the shape of the $C$ vs.\ $\theta$ curves remains the same. We have checked that the inferred values of $n_c$ and $h$ are distributed around their known values of $100$ and $0$, respectively, and that the shape persists across a range of filter window sizes.

These results suggest that the main findings above are not sensitive to our assumption that feedback is direct, and therefore that we are justified in using Eq.\ \ref{eq:rxns} as a coarse-grained model to analyze the T cell data.

\section{Discussion}
\label{sec:discussion}

We have employed the fact that a feedback-induced bifurcation exhibits the scaling properties of the mean-field Ising universality class to provide a simple prescription for modeling and analyzing biological data. Contrary to existing mixture-model approaches, our method is most valuable near the bifurcation point, which is where biologically significant cell-fate decisions are expected to take place. Our approach provides the effective order parameter, reduced temperature, magnetic field, and heat capacity from experimental distributions without fitting or needing to know the molecular details. By applying the approach to T cell flow cytometry data, we discovered that these quantities discriminate between cellular responses in an intuitive, interpretable way, and that the heat capacity allows estimation of the molecule number from fluorescence intensity for a bifurcating system. By generalizing the theory to include indirect feedback, we demonstrated the capacity to model realistic signaling cascades where indirect feedback is common. Our approach should be applicable to other systems observed to undergo a pitchfork-like bifurcation and the associated unimodal-to-bimodal transition in abundance distributions, but not to systems which have an absorbing or extinction state, as they are expected to fall under a different universality class \cite{ohtsuki1987nonequilibrium, grassberger1978reggeon}.

The theory assumes only birth-death reactions and neglects more complex mechanisms such as bursting \cite{friedman2006linking, mugler2009spectral} or parameter fluctuations \cite{shahrezaei2008colored, horsthemke1984noise}. These mechanisms are known to produce long tails and may be responsible for the long tails observed in the experimental data [Fig.\ \ref{fig:experiment}(a) and (b)]. Cell-to-cell variability (CCV) may also contribute to the long tails, as it is known to be present in T cell populations \cite{cotari2013cell}. Our theory neglects CCV and instead assumes that the distribution of molecule numbers across the population is the same as that traced out by a single cell over time. Although CCV may play an important role, one generically expects the role of intrinsic fluctuations to be amplified near a critical point, and models that ignore CCV have been shown to be sufficient to explain both the bimodality \cite{das2009digital} and variance properties \cite{prill2015noise} of ppERK in T cells. Moreover, the fact that our theory provides an estimate of the molecule number that is consistent with other estimates suggests that intrinsic fluctuations play a large role. Distinguishing between intrinsic fluctuations and long-lived CCV is an important topic for future work.

Our work provides key tools that can be used for a broader exploration of biological systems. The approach is applicable to any experimental dataset that exhibits unimodal and bimodal abundance distributions, and could lead to a unified picture of diverse cell types and environmental perturbations in terms of effective thermodynamic quantities. At the same time, several extensions of our work are natural. For example, the dynamics of the theory could be probed to investigate the consequences of critical slowing down for driven or dynamically perturbed systems with feedback. Alternatively, the theory could be generalized to systems that are not well-mixed, such as intracellular compartments or communicating populations, to investigate space-dependent universal behavior and its biological implications.

\section{Data availability}
Data and code for all figures and the MIFlowCyt record are available at\\
\url{https://github.com/AmirErez/UniversalImmune}.

\section*{Acknowledgments}
This work was supported by Human Frontier Science Program grant LT000123/2014 (Amir Erez), National Institutes of Health (NIH) grant R01 GM082938 (A.E.), Simons Foundation grant 376198 (T.A.B.\ and A.M.), and the Intramural Research Program of the NIH, Center for Cancer Research, National Cancer Institute.

\appendix

\section{Mapping for Schl\"ogl and Hill models}
\label{app:param}

Here we provide the mapping from $n_c$, $\theta$, and $h$ to the biochemical parameters and vice versa for the Schl\"ogl and Hill models. For the Schl\"ogl model, the feedback function is
\begin{equation}
f_n = \frac{aK^2+s(n-1)(n-2)}{(n-1)(n-2)+K^2}.
\end{equation}
The condition $f''_{n_c} = 0$ is satisfied by
\begin{equation}
n_c = \frac{3}{2} + \frac{1}{6}\sqrt{3(4K^2-1)}.
\end{equation}
The parameters $\theta$ and $h$ are given by Eq.\ \ref{eq:cparam}, where
\begin{align}
f_{n_c} &= \frac{(3a+s)K^2-s}{4K^2-1}, \\
f'_{n_c} &= (s-a)K^2\left(\frac{3}{4K^2-1}\right)^{3/2}, \\
f'''_{n_c} &= -6(s-a)K^2\left(\frac{3}{4K^2-1}\right)^{5/2}.
\end{align}
These expressions are inverted to write the biochemical parameters $a$, $s$, and $K$ in terms of $n_c$, $\theta$, and $h$:
\begin{align}
K^2 &= \frac{1}{4}(3x^2+1), \\
s &= \frac{3n_c^3(\theta+h) + n_cx^2 + x^3}{3n_c^2\theta + x^2}, \\
a &= \frac{(3x^2+1)[3n_c^3(\theta+h) + n_cx^2 + x^3] - 4x^5}{(3x^2+1)(3n_c^2\theta + x^2)},
\end{align}
where $x \equiv 2n_c - 3$.

Similarly, for the Hill model we have
\begin{align}
\label{eq:Hill}
f_n &= a + s\frac{n^H}{n^H+K^H}, \\
n_c &= K\left(\frac{H-1}{H+1}\right)^{1/H}, \\
f_{n_c} &= a + \left(\frac{H-1}{2H}\right)s, \\
f'_{n_c} &= \frac{(H^2-1)s}{4Hn_c}, \\
f'''_{n_c} &= -\frac{(H^2-1)^2s}{8Hn_c^3}, \\
K &= n_c\left(\frac{H+1}{H-1}\right)^{1/H}, \\
s &= n_c\frac{16H}{(H^2-1)[(H^2-1)\theta+4]}, \\
a &= n_c\frac{(H-1)[(H+1)^2(\theta+h)+4]}{(H+1)[(H^2-1)\theta+4]}.
\end{align}
In the Hill model, $H$ is an additional free parameter.

\section{Scaling exponents $\beta$, $\gamma$, and $\delta$}
\label{app:scalings}

\begin{figure}[t]
\centering
\includegraphics[width=\linewidth]{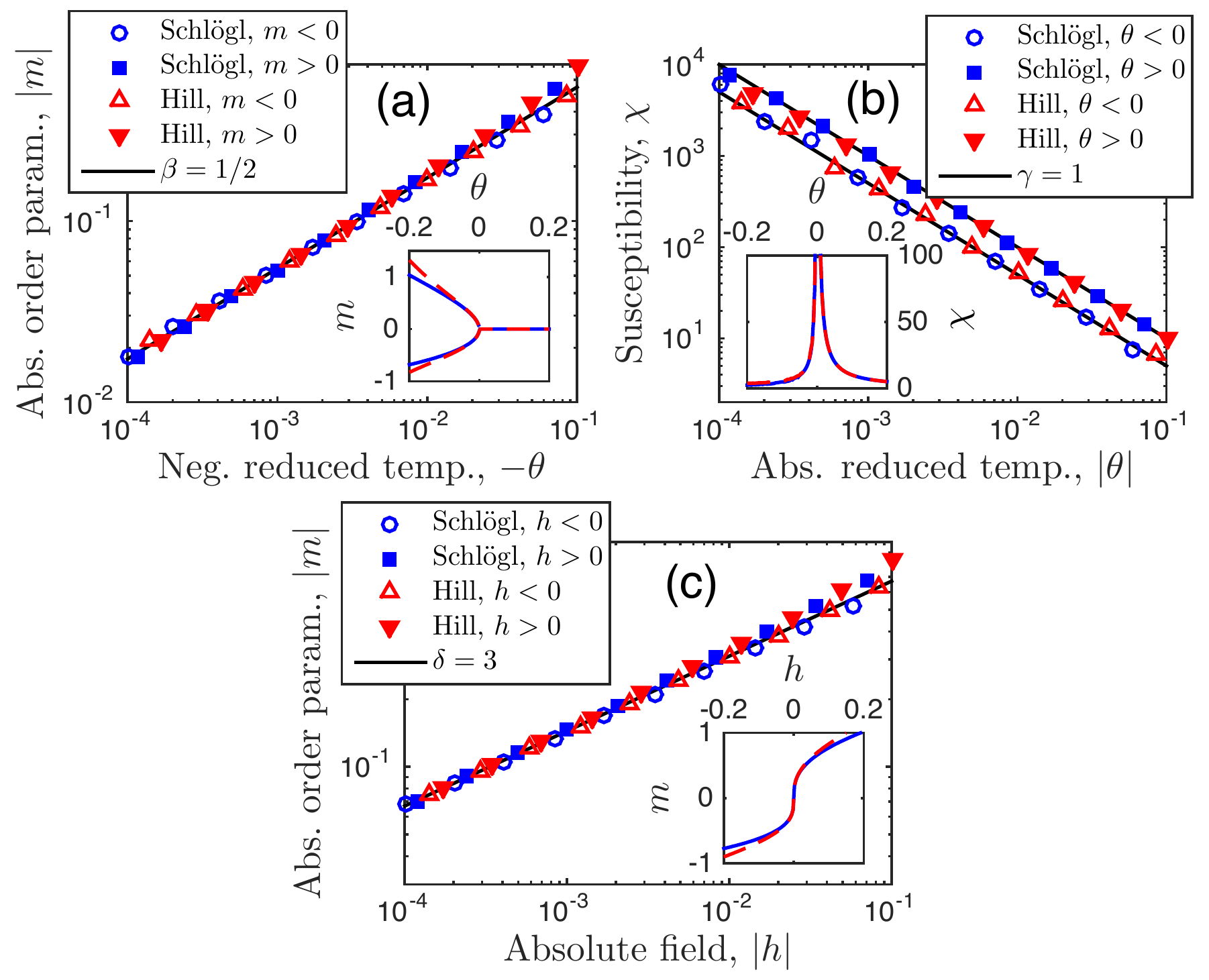}
\caption{Scaling exponents $\beta$, $\gamma$, and $\delta$ for biochemical feedback models agree with those of mean-field Ising universality class. Parameters: $H=3$ and $n_c = 500$.}
\label{fig:scalings}
\end{figure}

Here we verify that the stochastic Schl\"ogl and Hill models have the scaling exponents $\beta$, $\gamma$, and $\delta$ of the mean-field Ising universality class. Specifically, we expect $m=\pm(-3\theta)^\beta$ for $h=0$ and $\theta < 0$, with $\beta = 1/2$; $\chi = \theta^{-\gamma}$ or $\chi = (-2\theta)^{-\gamma}$ for $\theta > 0$ or $\theta < 0$, respectively, with $\gamma = 1$, where $\chi \equiv (\partial_h m)_{h=0}$ is the dimensionless susceptibility; and $m = (3h)^{1/\delta}$ for $\theta = 0$, with $\delta = 3$. Fig.\ \ref{fig:scalings} computes these quantities from the parameters and maxima of $p_n$ for the Schl\"ogl and Hill models using the mapping in Eq.\ \ref{eq:cparam}. We see that the scalings hold, as expected.

\section{Experimental methods}
\label{app:expt}

The experimental data analyzed in Fig.\ \ref{fig:experiment}, along with a detailed description of the experimental methods, have been published previously \cite{vogel2016dichotomy}. In this section we briefly summarize the experimental system and methods. The drugs and dose ranges used Figs.\ \ref{fig:experiment} and \ref{fig:heat2} are listed in Table \ref{tab:drugs}.

\begin{table}[b]
\begin{tabular}{|l|l|l|l|}
	\hline
	Drug & Inhibits & Dose range (nM) & Shape in Fig.\ \ref{fig:heat2} \\
	\hline
	\hline
	PD325901 & MEK & $0.09$$-$$1000$ & Up triangle \\
	\hline
	AZD6244 & MEK & $2.4$$-$$5000$ & Down triangle \\
	\hline
	Trametinib & MEK & $0.5$$-$$1000$ & Right triangle \\
	\hline
	Dasatinib & SRC & $0.09$$-$$1000$ & Circle \\
	\hline
	Bosutinib & SRC & $0.5$$-$$1000$ & Square \\
	\hline
	PP2 & SRC & $24$$-$$50,$$000$ & Diamond \\
	\hline
\end{tabular}
\caption{Drugs and dose ranges of experimental data \cite{vogel2016dichotomy}. Doses are spaced logarithmically. Fig.\ \ref{fig:experiment} uses PD325901 and Dasatinib. Fig.\ \ref{fig:heat2} uses all drugs.}
\label{tab:drugs}
\end{table}

The data investigate inhibition of the antigen-driven MAP kinase cascade in primary CD8+ mouse T cells. A natural way to stimulate T cells is to load a peptide (a fragment of an antigenic protein that the T cells are programmed to recognize) onto antigen-presenting cells. We achieve this by incubating RMA-S cells with antigen at $37$ $^{\rm o}$C. At the same time, we harvest the spleen and lymph nodes of a RAG2$^{-/-}$ OT1 mouse which has T cells specific only to the ovalbumin peptide with the amino acid sequence SIINFEKL. When we mix the OT1 T cells with the antigen-loaded RMA-S cells, we expose the OT1 T cells to their activating peptide. In response, the T cells activate their receptors through a SRC Family kinase (Lck). This triggers an enzymatic cascade, which in turn actives Ras-Raf-MEK-ERK leading to double phosphorylation of ERK, rendering it capable of communicating with the nucleus. By waiting for 10 minutes, the signaling reaches steady state and the distribution of the abundance of doubly phosphorylated ERK (ppERK) is the readout.

To measure the abundance of ppERK, we use fluorescence cytometry. Specifically, we introduce ppERK-targeted antibodies that are pre-conjugated with a fluorescent dye. Because antibodies selectively attach to their target molecule with negligible false-positives, the fluorescence intensity of the dye is proportional to the abundance of ppERK. To measure the intensity, approximately 30,000 cells per sample are passed one-by-one through a microfluidic device where they encounter a series of excitation lasers. Each cell yields one intensity value, and the histogram provides an estimate of the distribution of ppERK abundance across the population. We assume that the distribution across the population is a fair representation of the steady-state distribution of ppERK abundance of a single cell. This is reasonable (and is the accepted practice) since while the cells are alive and the experiment is taking place, they are in a dilute suspension (approximately 30,000 cells in 100 $\mu$L), not close enough together to influence each other.

\section{Experimental data analysis}
\label{app:analysis}

We calculate the forcing functions and the effective thermodynamic quantities $I_c$, $\theta$, $h$, and $C$ from an experimental intensity distribution using the following procedure. First, we set $n = I/I_1$ to convert $p(I)$ to $p_n$, where the intensity of one molecule $I_1$ converts from intensity $I$ to molecule number $n$. We will see below that only $C$ will depend on the value of $I_1$.

Next, because the experimental distributions are long-tailed, we convert Eq.\ \ref{eq:cparam} to $\log I$ space for numerical stability. Here we provide the necessary conversions between functions of $n$ from the theory, and functions of $\ell \equiv \log I$ from the experiments, as the probability distributions over $n$ and $\ell$ do not have the same functional forms \cite{erez2018modeling}. In what follows, prime denotes the derivative of a function with respect to its argument ($n$ for $f$; and $\ell$ for $q$, $Q$, and $\phi$). $\ell$ and $n$ are related as
\begin{equation}
\label{eq:var}
\ell = \log(I_1n), \qquad
n = \frac{e^\ell}{I_1}.
\end{equation}
We denote the distribution of $\ell$ as $q(\ell)$. Approximating $n$ as continuous, probability conservation requires
\begin{equation}
\label{eq:q}
q(\ell) = \frac{p_n}{d\ell/dn} = np_n.
\end{equation}
Using Eq.\ \ref{eq:q}, the feedback function (Eq.\ \ref{eq:fn}) is
\begin{equation}
\label{eq:f}
f_n = \frac{np_n}{p_{n-1}}
	= (n-1)\frac{np_n}{(n-1)p_{n-1}}
	= (n-1)\frac{q(\ell)}{q(\tilde{\ell})},
\end{equation}
where, using Eq.\ \ref{eq:var},
\begin{equation}
\tilde{\ell} = \log[I_1(n-1)]
	= \log(I_1n) + \log(1-\epsilon)
	\approx \ell -\epsilon.
\end{equation}
The last steps define $\epsilon \equiv 1/n$ and assume that for most values of $n$ with appreciable probability we have $\epsilon \ll 1$. Therefore Eq.\ \ref{eq:f} becomes
\begin{align}
f_n &= n(1-\epsilon)\frac{q(\ell)}{q(\ell-\epsilon)}
	\approx n(1-\epsilon)\frac{q(\ell)}{q(\ell)-\epsilon q'(\ell)}
	= \frac{n(1-\epsilon)}{1-\epsilon q'/q} \nonumber \\
	&\approx n(1-\epsilon)\left(1+\epsilon \frac{q'}{q}\right)
	\approx n\left[1+\epsilon\left(\frac{q'}{q}-1\right)\right] \nonumber \\
\label{eq:f2}
&= n + \frac{q'}{q} - 1,
\end{align}
where we have kept to first order in $\epsilon$. Defining $\phi(\ell) \equiv f_n - n$, from Eq.\ \ref{eq:f2} we have
\begin{equation}
\label{eq:phi}
\phi(\ell) = f_n - n = \frac{q'}{q} - 1 = Q'.
\end{equation}
In the last step we define $Q(\ell) \equiv -\ell + \log q$ so that $\phi$ is computed as a total derivative, which we find more numerically stable. The $\phi(\ell)$ are the forcing functions plotted in Fig.\ \ref{fig:experiment}(c) and (d).

The point $n_c$ is defined by $f''_{n_c} = 0$. Eq.\ \ref{eq:var} implies
\begin{equation}
\label{eq:deriv}
\partial_n = I_1e^{-\ell}\partial_\ell,
\end{equation}
such that the condition $f''_{n_c} = 0$ becomes
\begin{align}
0 &= \partial_n^2f
	= \partial_n^2(\phi+n)
	= \partial_n^2\phi
	= (I_1e^{-\ell}\partial_\ell)^2\phi \nonumber\\
	&= I_1e^{-\ell}\partial_\ell (I_1e^{-\ell}\partial_\ell\phi)
	= I_1^2e^{-\ell}\left(-e^{-\ell}\phi' + e^{-\ell}\phi''\right) \nonumber\\
	&= (I_1^2e^{-\ell})^2\left(\phi'' - \phi'\right).
\end{align}
Therefore, we define a point $\ell_c$ by
\begin{equation}
\label{eq:lc}
\phi''(\ell_c) = \phi'(\ell_c).
\end{equation}
Numerically we enforce Eq.\ \ref{eq:lc} by writing it as $0 = \partial_\ell\left(\phi'-\phi\right)$, and therefore
\begin{equation}
\label{eq:lc2}
\ell_c = {\rm argmax}_\ell \left(\phi'-\phi\right).
\end{equation}
Then
\begin{equation}
\label{eq:nc}
I_c = e^{\ell_c}, \qquad n_c = \frac{e^{\ell_c}}{I_1}
\end{equation}
from Eq.\ \ref{eq:var}.

Derivatives of $f$ with respect to $n$ at $n_c$ are related in a straightforward way to derivatives of $\phi$ with respect to $\ell$ at $\ell_c$. First, the zeroth derivative is, by Eq.\ \ref{eq:phi},
\begin{equation}
\label{eq:f0}
f_{n_c} = \phi(\ell_c) + n_c,
\end{equation}
where $n_c$ is defined in Eq.\ \ref{eq:nc}. Then, using Eq.\ \ref{eq:deriv}, the first derivative is
\begin{align}
f'_{n_c} &= \partial_n[\phi+n]_{n_c}
	= \partial_n[\phi]_{n_c} + 1
	= \left[(I_1e^{-\ell}\partial_\ell)\phi\right]_{\ell_c} + 1 \nonumber\\
\label{eq:f1}
	&= I_1e^{-\ell_c}\phi'(\ell_c) + 1
	= \frac{\phi'(\ell_c)}{n_c} + 1.
\end{align}
Finally, by a similar procedure, the third derivative is
\begin{align}
\label{eq:f3}
f'''_{n_c} &= \frac{1}{n_c^3}\left[\phi'''(\ell_c) - 3\phi''(\ell_c) + 2\phi'(\ell_c)\right] \nonumber\\
	&= \frac{1}{n_c^3}\left[\phi'''(\ell_c) - \phi'(\ell_c)\right],
\end{align}
where the second step uses Eq.\ \ref{eq:lc}.
Using Eqs.\ \ref{eq:nc}-\ref{eq:f3}, $\theta$ and $h$ (Eq.\ \ref{eq:cparam}) become
\begin{align}
\label{eq:theta}
\theta &= \frac{2(1-f'_{n_c})}{-f'''_{n_c}n_c^2}
	= \frac{-2\phi'(\ell_c)}{\phi'(\ell_c) - \phi'''(\ell_c)}, \\
\label{eq:h}
h &= \frac{2(f_{n_c} - n_c)}{-f'''_{n_c}n_c^3}
	= \frac{2\phi(\ell_c)}{\phi'(\ell_c) - \phi'''(\ell_c)}.
\end{align}
Note that they do not depend on $I_1$.

To estimate the derivatives in Eqs.\ \ref{eq:theta} and \ref{eq:h}, we apply a Savitsky-Golay filter to the experimental $q(\ell)$ \cite{savitzky1964smoothing}. Savitsky-Golay filtering replaces each data point with the value of a polynomial of order $J$ that is fit to the data within a window $W$ of the point. Since we require three derivatives of $\phi(\ell)$ (Eqs.\ \ref{eq:theta} and \ref{eq:h}), which depends on the first derivative of $q(\ell)$ (Eq.\ \ref{eq:phi}), we use the minimum value $J=4$. Thus, the procedure requires the adjustable parameter $W/L$, where $L$ is the number of $\log I$ bins. We find that $L=100$ and $W = 25$ suffice [Figs.\ \ref{fig:experiment}, \ref{fig:heat2}, and \ref{fig:indirect}(b)], and that results are robust to $W/L$.

\begin{figure}[t]
\centering
\includegraphics[width=\linewidth]{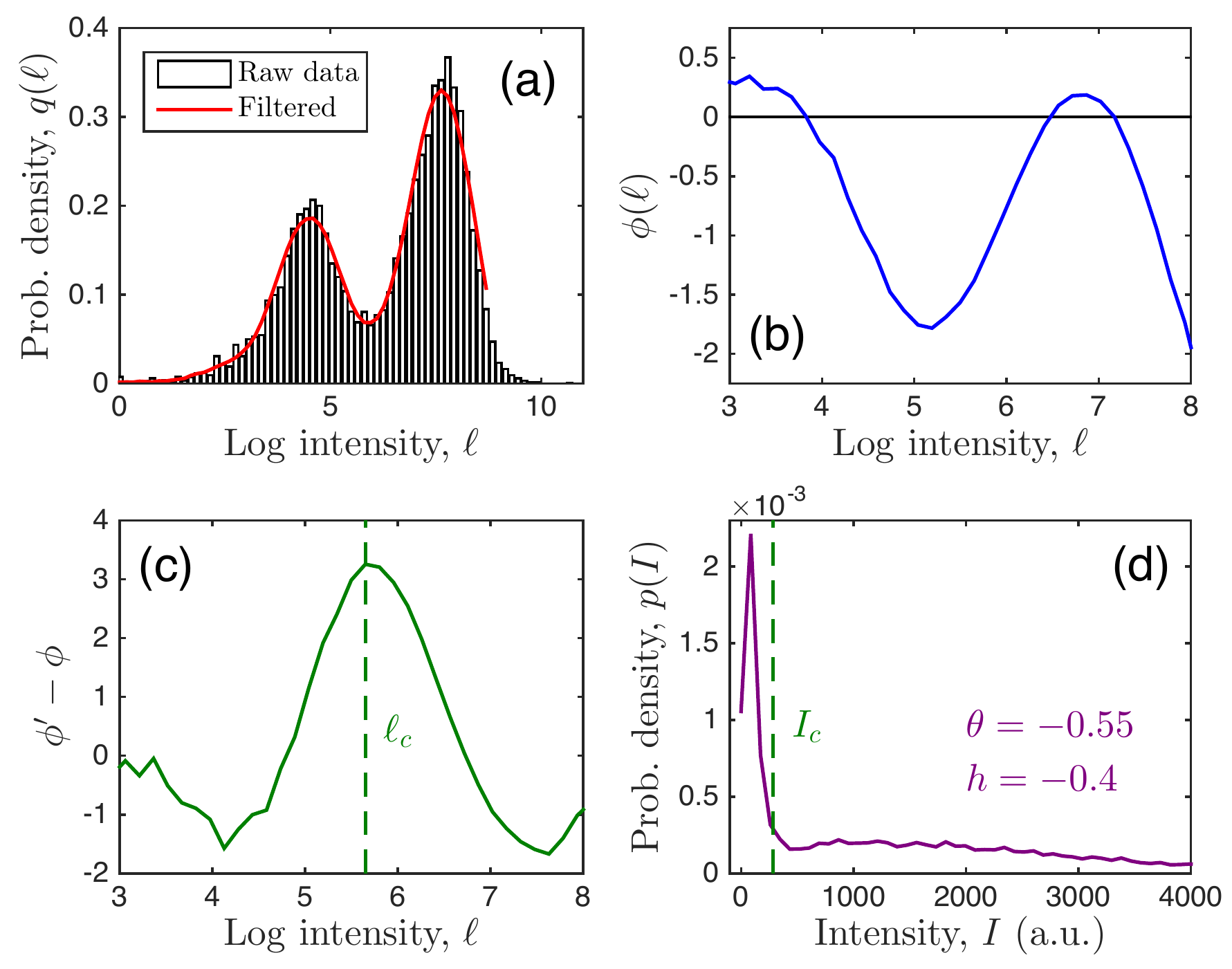}
\caption{Demonstration of analysis procedure for $3.4$ nM of MEK inhibitor PD325901. Parameters: $L=100$ and $W=25$.}
\label{fig:analysis}
\end{figure}

The analysis is demonstrated for an example experimental distribution in Fig.\ \ref{fig:analysis}. In summary, we:
\begin{enumerate}
\item plot $q(\ell)$ from the data using $L$ bins [Fig.\ \ref{fig:analysis}(a), black];
\item filter $q(\ell)$ using window $W$ [Fig.\ \ref{fig:analysis}(a), red];
\item compute $\phi(\ell)$ using Eq.\ \ref{eq:phi} [Fig.\ \ref{fig:analysis}(b)];
\item compute $\ell_c$ using Eq.\ \ref{eq:lc2} [Fig.\ \ref{fig:analysis}(c)];
\item compute $I_c$, $\theta$, and $h$ from $\ell_c$, $\phi$, and its derivatives using Eqs.\ \ref{eq:nc}, \ref{eq:theta}, and \ref{eq:h};
\item compute $p_n$ from the data using $I_1$; and
\item compute $C/k_{\rm B}$ from $p_n$, $\theta$, $n_c$ (Eq.\ \ref{eq:nc}), $f'''_{n_c}$ (Eq.\ \ref{eq:f3}), and $f_n$ (Eq.\ \ref{eq:f2}) using Eq.\ \ref{eq:C}.
\end{enumerate}
Fig.\ \ref{fig:analysis}(d) shows that $I_c$ falls between the maxima as expected, and that $\theta$ and $h$ are negative corresponding to a distribution that is bimodal and skewed to the left, respectively.

To estimate the value of $I_1$, consider $\chi^2$, defined as
\begin{equation}
\label{eq:SSE}
\chi^2 = \sum_{i=1}^N \frac{1}{\sigma_i^2}\left[\frac{C_i}{k_{\rm B}} - \frac{C(\theta_i)}{k_{\rm B}}\right]^2,
\end{equation}
where $N$ is the number of data points, $C_i/k_{\rm B}$ is the value of the heat capacity for each data point, $C(\theta_i)/k_{\rm B}$ is the predicted value of the heat capacity at the location $\theta_i$ of that data point, and $\sigma_i^2$ is the variance for data point $i$. Under the simplifying assumption that $\sigma_i^2$ takes the same value $\sigma^2$ for all data points, we have $\chi^2 = s/\sigma^2$, where $s$ is the sum of squared errors plotted in Fig.\ \ref{fig:heat2}(e). As a function of $I_1$, $\chi^2$ should scale quadratically near its minimum,
\begin{equation}
\chi^2 = \frac{(I_1-\bar{I}_1)^2}{\sigma_{I_1}^2} + {\rm const},
\end{equation}
where the location and curvature of the minimum give the best estimate $\bar{I}_1$ and error in the estimate $\sigma_{I_1}$, respectively \cite{bevington2002data}. In terms of $s$ we have
\begin{equation}
\label{eq:s}
s = \sigma^2\frac{(I_1-\bar{I}_1)^2}{\sigma_{I_1}^2} + s^*,
\end{equation}
where $s^*$ is the minimal value.
The value $\sigma^2$ is, by definition, the average squared deviation of the data from the theory \cite{bevington2002data},
\begin{equation}
\sigma^2 = \frac{1}{N} \sum_{i=1}^N \left[\frac{C_i}{k_{\rm B}} - \frac{C(\theta_i)}{k_{\rm B}}\right]^2
	= \frac{s^*}{N},
\end{equation}
here evaluated at the minimum $s^*$. Inserting this result into Eq.\ \ref{eq:s}, we obtain
\begin{equation}
s = s^*\left[\frac{(I_1-\bar{I}_1)^2}{N\sigma_{I_1}^2} + 1\right].
\end{equation}
We see that if $I_1$ deviates from $\bar{I}_1$ by $\sigma_{I_1}$, then $s$ is larger than its minimal value by a factor of $1 + N^{-1}$. This criterion, illustrated by the black line in Fig.\ \ref{fig:heat2}(e), is used to determine $\sigma_{I_1}$.

Fig.\ \ref{fig:heat2}(e) is restricted to data whose $\theta$ values are less than or equal to $\Delta\theta = 0.05$ in magnitude, of which there are $N = 20$ points. As $\Delta\theta$ increases, $N$ increases, and the minimum of $s$ also becomes less sharp. These effects compensate, yielding an estimate of $I_1$ whose value and error are insensitive to $\Delta\theta$, as seen in Fig.\ \ref{fig:heat2}(f). Averaged across $\Delta\theta$ values, we find $I_1 = 0.5$ and $\sigma_{I_1} = 0.2$, as reported in the main text.

We compare our estimate of ppERK molecule number to two previous studies. In \cite{altan2005modeling}, it was estimated that there are 100,000 ERK molecules per cell (see Results in \cite{altan2005modeling}). From \cite{hukelmann2016cytotoxic}, we estimate that there are 214,000 ERK molecules per cell. Specifically, from the Excel file associated with Fig.\ 1 in \cite{hukelmann2016cytotoxic}, we sum the mean number (column I) of ERK1 (also called MAPK3, row 2345) and ERK2 (also called MAPK1, row 874) to obtain 214,000 molecules to three significant digits. In \cite{altan2005modeling}, it was estimated that 50\% of ERK molecules are doubly phosphorylated during T cell receptor activation (see caption of Fig.\ S2 in \cite{altan2005modeling}).


\end{document}